\documentclass[prl,reprint]{revtex4-1}
\usepackage{graphicx}
\usepackage{amsmath}
\usepackage[dvipsnames,usenames]{color}

\begin{document}
\title{Zeta potential dependent Self-electrophoresis of Pt-coated Janus particles in hydrogen peroxide solutions}
\author{Cheng-Tse Wu}

\author{Yen-Wei Lin}
\author{Hong-Ren Jiang}
\email{hrjiang@iam.ntu.edu.tw}
\affiliation{Institute of Applied Mechanics, National Taiwan University, No.1, Sec.4, Roosevelt Rd., Taipei 10617, Taiwan}

\begin{abstract}
We provide experimental results to show that self-propulsion of Janus particles made by coating platinum on the hemisphere of dielectric particles in hydrogen peroxide solution is similar to self-electrophoresis. By different surface treatments and measuring the motion of particles and their $\zeta$-potentials, we find that the speed and direction of motion are determined by the $\zeta$-potential in a given concentration of hydrogen peroxide solution. When sign of $\zeta$-potential is changed from negative to positive, the direction of motion reverses from toward non-catalytic side to catalytic side. We also find that the angular distribution of Janus particle is more polarized with increasing of the concentration of hydrogen peroxide, which support the self-electrophoresis mechanism.
\end{abstract}

\maketitle

  Macro-scale engines can be easily achieved by charging chemical fuels which are converted into kinetic energy. However, to design micro-scale engines, it is important to know how to distribute energy in micro-scale systems, how to design mechanism and how to control the motion with predicted functions. For coupling energy dissipation and motion, phoretic transports are widely used to move, and even control particles in micro-scale systems by applying external fields such as concentration gradient, electric field or thermal gradient[\textcolor{Blue}{1}-\textcolor{Blue}{3}]. Phoretic mechanisms are recently used to design self-propelling objects by arising self-generated non-equilibrium conditions[\textcolor{Blue}{4}-\textcolor{Blue}{15}], that is colloids which can create asymmetric forces or asymmetric surrounding condition would enable active motion.

For utilizing chemical reaction and phoretic mechanism for self-propulsion of colloids, self-diffusiophoresis has been proposed, where chemical reaction occurring on a particle can generate a local concentration gradient and the particle can move along the self-generated gradient[\textcolor{Blue}{5},\textcolor{Blue}{7}-\textcolor{Blue}{12},\textcolor{Blue}{14}-\textcolor{Blue}{15}]. In experiments, several groups observed that a metal-dielectric Janus particle made by coating platinum on its hemisphere in hydrogen peroxide solution converts chemical energy to kinetic energy and causes self-propulsion[\textcolor{Blue}{6}]. However, the direction of this kind of self-propulsion can both toward the noncoated (non-catalytic) side or coated (catalytic) side[\textcolor{Blue}{7},\textcolor{Blue}{8}], which rises discussion about the mechanism of the motion[\textcolor{Blue}{9}-\textcolor{Blue}{15}]. Regarding the mechanism, besides self-diffusiophoresis, bubble propulsion model[\textcolor{Blue}{13}] and self-electrophoresis[\textcolor{Blue}{14},\textcolor{Blue}{15}] are also proposed. In the bubble propulsion model, bubbles occur and leave on the Pt surface and a particle would move toward the non-coating side due to conservation of momentum. In the self-electrophoresis, current generated due to uneven surface chemical reactions, which similar in the bi-metal swimmers, are proposed. However, due to lacking of direct measurements of local physical parameters involved in proposed mechanisms, the mechanism is still not well understood.

In this study, we experimentally modify and measure the surfaces of Pt-silica Janus particles to find the relation between their motion and their surface charges. We find the moving direction, whether move to the coated side or noncoated side, indeed dependent on the $\zeta$-potential of the particle, which can be regulated by both of  chemical modification and surfactant adsorption. Simple schematic figures of self-propulsion of Janus particle with original negative $\zeta$-potential and positive $\zeta$-potential obtained by chemical modification of aminosilane-APTES are shown in Fig.1.(a)(b). In our experiments, a Janus particle with a negative $\zeta$-potential moves toward the non-coating side and a Janus particle with a positive $\zeta$-potential moves toward the coating side. 

\begin{figure}[h]
\begin{center}
\includegraphics[width=8cm]{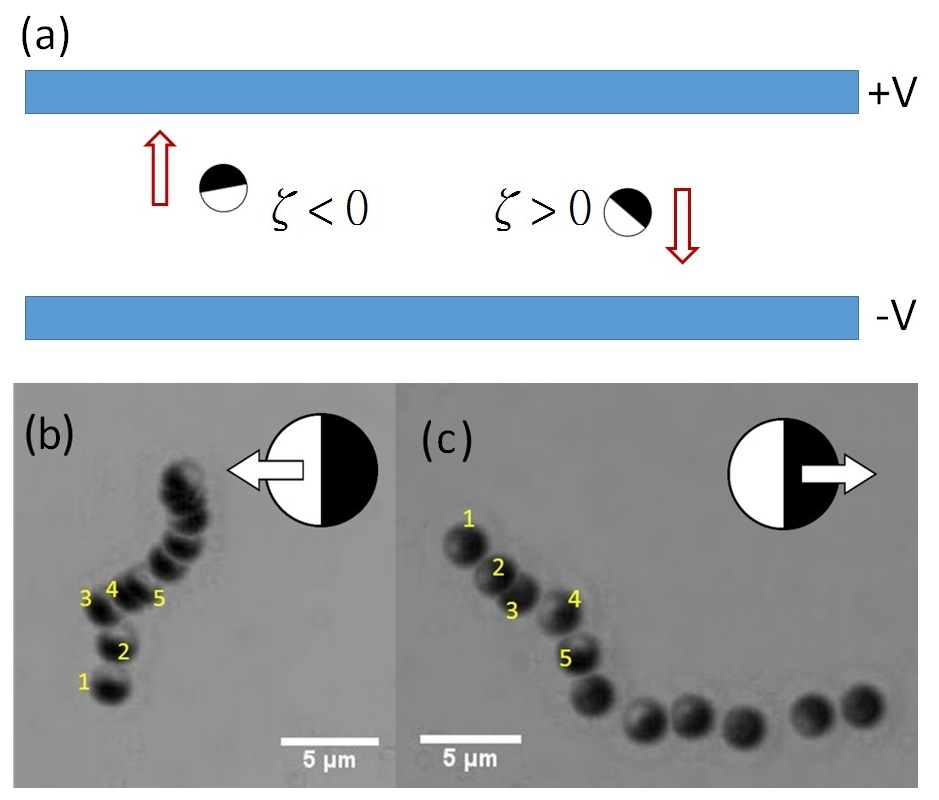}
\caption{ (a) We built a hand-made device to detect the $\zeta$-potential of individual colloidal particle in solution. As $\zeta$$<$0, particle goes to positively charged plate, and vice versa. (b)(c) The Janus particles in hydrogen peroxide shows opposite self-propelling motion in different $\zeta$-potential surface conditions. To determine the directions of JP swimmer, we number the particles as time sequential order from 1 to 5. (b)$\zeta$$<$0; (c)$\zeta$$>$0.}           

\label{induction loop} 
\end{center}
\end{figure}
To prepare Janus particles, a monolayer of 2 $\mu$m silica particles is prepared by a drying process. Thin layer of platinum is deposited by DC sputter coating to create $\sim$30 nm thick coating on the half side of each particles.
The coating of a particle can be visualized under an optical microscope as shown in Fig.1 ,and therefore the direction of motion can be directly measured. The dark side of a particle has Pt coating, which does not transmit light and thus darker than the other side (silica)[\textcolor{Blue}{7}]. A thin chamber containing a solution and particles is sandwiched by two cover glasses with a $\sim$100 $\mu$m spacer. Images of particles during self-propulsion are collected through an objective (\textit{40X NA 0.55}) by a camera at 20 frames per second. The positions of a particle in different frames are determined by images through an particle tracing program.
For the experimental study of adsorption on particle surface, all surfactants were purchased from Tokyo Chemical Industrial Japan. The H$_2$O$_2$ solution was diluted to certain concentration mixed homogeneously with Janus particles and ions or surfactants. We operated the surfactants in H$_2$O$_2$ solution in the unit of CMC (critical micelle concentration). The critical micelle concentration is 0.662 \% (w/w) to DDAB, 1.34 \% (w/w) to STAC, 0.07 \% (w/w) to TWEEN 20, 0.23 \% (w/w) to SDS [\textcolor{Blue}{16}-\textcolor{Blue}{19}]. 

The $\zeta$-potential is not directly measurable but can be calculated by experimentally-determined electrophoretic mobility[\textcolor{Blue}{20}]. Commercial $\zeta$-potential measuring instrument \textit{Malvern Zetasizer Nano} basing on \textit{M3-PALS (Phase analysis Light Scattering)} is dedicated to the measurement of $\zeta$-potential , which measures the mean mobilities by using microelecrophresis in capillary cell. To measure the individual electric charge of single particle, we built a simple device which two ITO glasses sandwiching solution of colloidal particles. As a DC electric field is vertically applied by ITO glasses, the particles move vertically, and this motion can be detected by setting CMOS camera to observe the the change of diffraction ring. By converting the change rate of diffraction ring to drift velocity, we can obtain the real charge of individual Janus particle. We measured the velocities of particles in different concentrations of hydrogen peroxide solutions to characterize the general features of the motion. The velocity of the direct motion is analyzed by two parameters fitting method within the rotational diffusion time scale ($\sim$20 s) for  2$\mu$m particles. The mean square displacement is given as $\Delta L^{2}=4D\Delta t+ V^{2}\Delta t^{2}$ while $\Delta t$ is much smaller than rotational diffusion time scale. The diffusion coefficient $D$ and self-propelled velocity $V$ can be obtained by this method[\textcolor{Blue}{6}]. We observed that  particles move faster in the higher concentration conditions and the direction of motion follows the polarity of the Janus particle, toward the noncoated side (See movie-1 in Supplementary Information(SI)). Similar results have been shown in the previous studies[\textcolor{Blue}{7}].

We suspect that the self-propelling motion of Pt-coated Janus particles in H$_2$O$_2$ is strongly related to the surface charge. The self-propelling motion of Janus particles coated by aminosilane-APTES goes reverse in H$_2$O$_2$ solution.
Since from the silane-coating we can only see qualitative result, to quantify the surface charge, we added different concentrations of surfactant to regulate the surface charge, and we introduce $\zeta$-potential, which is a good surface charge indicator.

Since the adsorption of surfactants would provide a systematic method to control the surface properties, we measure the self-prpelling motions of Pt-silica Janus particles in anionic surfactant SDS, non-ionic surfactant TWEEN 20, and cationic surfactant DDAB. All results of measured velocities of particles in 0.75$\%$ hydrogen peroxide solutions containing different kinds surfactants are shown in Fig.2.  As added anionic surfactant, SDS  concentration increases, the self-propelling velocity of Janus particles shows slight decrease. Similar result is also obtained in non-ionic surfactant, TWEEN 20, solution. However, motion of particles shows dramatically changes in the cationic surfactant, DDAB, solution. In experiment we found that the reversal of self-propelling motion happened as DDAB concentration $10^{-3 }$ to $10^{-2}$ CMC. Moreover, Janus particles move faster in H$_2$O$_2$ solution with DDAB concentration 0.005 CMC and 0.01 CMC than in the solution without adding any surfactant. In the case of DDAB  concentration 0.005 CMC, it shows that the self-propelling motion of Janus particles can be more than two times faster than motion without adding surfactants. The effect of added SDS has been investigated in previous studies[\textcolor{Blue}{21}], and the decrease of velocity has been attributed to the effect of surface tension changing. However, the hypothesis of surface tension cannot explain the phenomenon which small amount of DDAB in the solution causes the direction of self-propulsion to reverse.

\begin{figure}[h]
\begin{center}
\includegraphics[width=8cm]{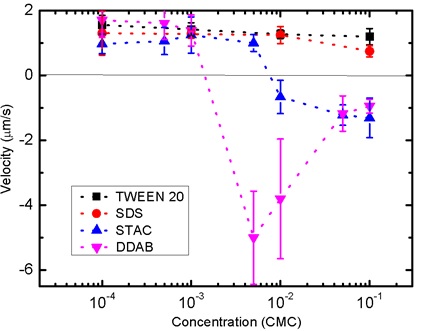}
\caption{The self-propelling velocity change in different kinds of surfactant  with 0.75\% H$_2$O$_2$. As shown in plot, non-ionic TWEEN 20 and anionic SDS slightly decrease the self-propelling velocity, in the other hand, cationic DDAB and STAC strongly change the behavior of self-propelling motion. Note that concentrations of the surfactants are normalized with critical micelles concentration (CMC).}           

\label{induction loop} 
\end{center}
\end{figure}

Since the change of surface tension cannot contribute to the reverse of self-propelling motion, our result shows that the effect of added cationic surfactant more likely comes from the modification of particle surface[\textcolor{Blue}{22}] instead of surface tension change. To conform the effect of cationic surfactants, we also measured motion of particles in another cationic surfactant, STAC, and obtained similar reversal behavior. From our results, both directly surface modification with positive charge silane or adding cationic surfactants lead the reversal of the direction of motion. The modification by silane can give the qualitative result, however, to obtain the quantitative evidence, we observed the effect of verifying the surface charge conditions. By adding cationic surfactants, DDAB and STAC, we can observe the phenomenon of self-propulsion motion changing while adding different concentration of surfactant. We measured $\zeta$-potential of Janus particles soaked in different concentrations of DDAB aqueous solution by both M3-PALS and self-built ITO-sandwiched device. Compare to the $\zeta$-potential measured by self-built ITO-sandwiched device, it shows positively shifted $\zeta$-potential by using M3-PALS measurement. The difference may come from that $\zeta$-potential measured by the M3-PALS is primarily coming from surfactant micelles in the solution. In Fig.3. is $\zeta$-potential of the Janus particle measured by self-built ITO-sandwiched device in different concentrations of STAC. The phenomenon that velocity of electrophretic motions reversed shows significant correspondence to the result of self-propelling motion of Janus particles, which we suggested a self-electrophretic model to it.  

\begin{figure}[h]
\begin{center}
\includegraphics[width=9cm]{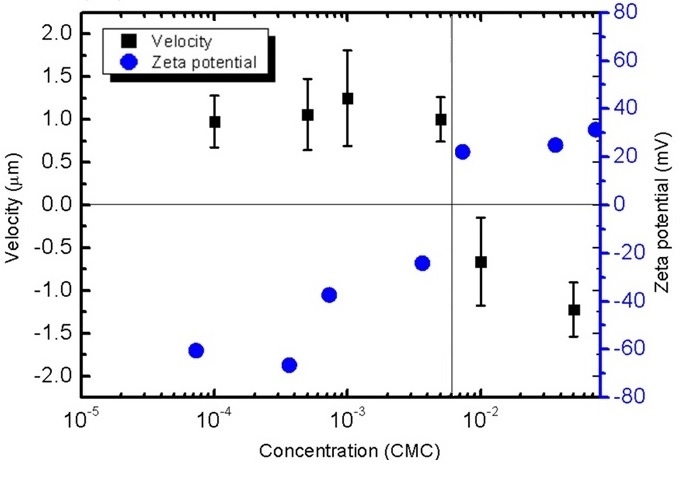}
\caption{The black dots show self-propelling velocity in different concentration of cationic surfactants STAC, and the blue dots show $\zeta$-potential measured by observing diffraction ring change rate in sandwiched ITO.}           

\label{induction loop} 
\end{center}
\end{figure}

The amount of surface charges is the main factor responding to the non-equilibrium condition created by Pt-catalyzed chemical reactions. Considering that the self-propulsion of Pt-coated Janus particle may be caused by a phoretic transport phenomenon, the most possible mechanism should be self-electrophoresis, where the motion is oriented towards the electric field generated by itself. And most likely the non-equilibrium condition created by Pt-catalyzed chemical reactions results from local electric field generated by itself, which effect only bulk liquid fluid near the surface. In fact, the effect of surface properties in reversal of direction of self-propulsion due to self-phoretic motion has been demonstrated in self-thermophoresis[\textcolor{Blue}{3}].

To visualize the flow field near particle surface associated with the motion of Janus particles, we stuck heavy Pt-coated Janus particles (8 $\mu$m) on the glass plate by drying out the Janus particles colloidal solution, then dropped H$_2$O$_2$ solution with small silica particles (2 $\mu$m). The liquid flow can be visualized by and observing the motion of small silica particles as tracers in the solution. In Fig.4.(a)(b) it shows that the heavy Janus particle which sinks on the bottom of liquid causes water flow nearby. To estimate the flow rate near Pt/silica interface, we coated Pt on glass slide with the same parameter as we coated the Janus particles, then H$_2$O$_2$ solution with silica particle tracers is laid on the glass slide. As the chemical reaction occurs, the silica particle tracers move from uncoated area toward the interface of Pt/glass. The velocity of tracers is calculated in different concentration as shown in Fig.4.(c)-(f) The more concentrated H$_2$O$_2$ solution is, the faster those particle tracers move toward Pt-coated surface. This indicates that there exists a fluid flow relating to the reaction of H$_2$O$_2$.

\begin{figure}[h]
\begin{center}
\includegraphics[width=8.5cm]{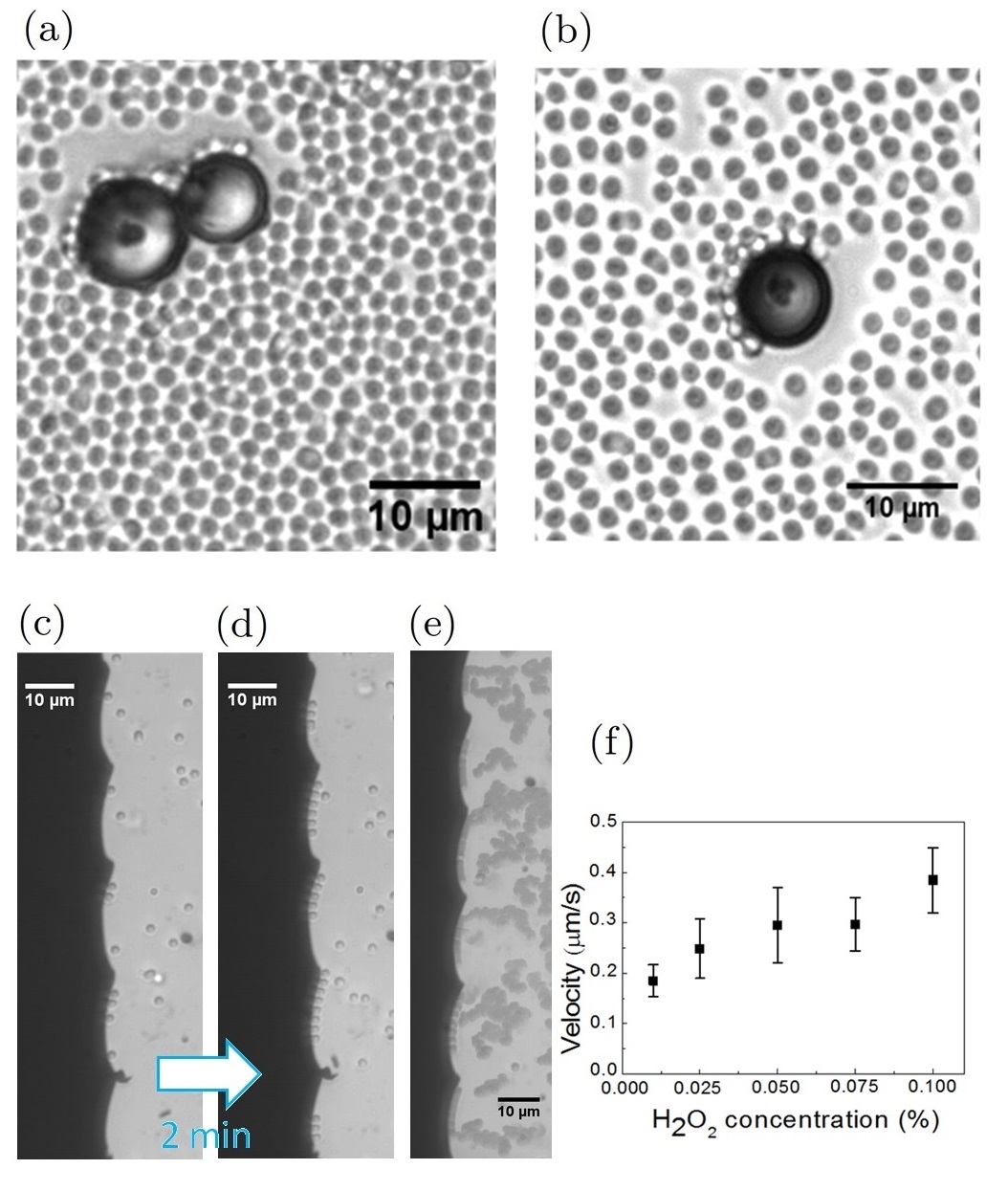}
\caption{(a)(b) 8$\mu$m Pt-coated Janus particles in H$_2$O$_2$ solution create liquid flow visualized by 2$\mu$m silica particle tracers. (a)Without adding DDAB; (b)With adding DDAB. Figure(c)(d) show that particle tracers move in 120 sec; (e)Stack of the particle tracers; (f)Particle tracers velocity in different concentration of H$_2$O$_2$ solution.}           

\label{induction loop} 
\end{center}
\end{figure}

First, we discuss the possible mechanism about how chemical reaction can create a steady electric field. The decomposition of H$_2$O$_2$ generating H$_2$O and O$_2$ is a heterogeneous catalytic reaction which occurs on the Pt surface, electrons and protons are separated from molecules and recombine continuously[\textcolor{Blue}{14},\textcolor{Blue}{23},\textcolor{Blue}{24}]. Note that protons can freely diffuse into the solution while electrons must stay on the Pt surface. Since the reaction occurs on the Pt surface, for continuous reaction, the supply of protons would be limited by diffusion but not for electrons. The limit of proton supply by diffusion allows more electrons always on the surface and then creates electric field which attracts protons to the surface. Thus steady electric field would be generated by the chemical reaction. The self-generated electric field applies force to the bulk fluid which contains opposite sign of charge to the surface charge, it causes Janus particles move away the coated side as $\zeta$$<$0 and toward the coated side as $\zeta$$>$0. The velocity field around the particle $v$ can be obtained by solving the Stokes equation $\eta \bigtriangledown^{2} v= \bigtriangledown p-f$. For a lager particle, the flow field is the same as classical electro-osmosis treated by Smoluchowski. The relation between propulsive velocity $U_{self-ep}$ for a self-electrophoresis and self-generated electric field $E^*$ can be shown as $U_{self-ep}=\frac{\varepsilon\varepsilon_0}{\eta }\zeta E^*$ ,where $\varepsilon$, $\varepsilon_0$, $ \eta$ are relative permittivity of fluid, vacuum permittivity, and $\zeta$-potential of the particle.

\begin{figure}[h]
\begin{center}
\includegraphics[width=9.5cm]{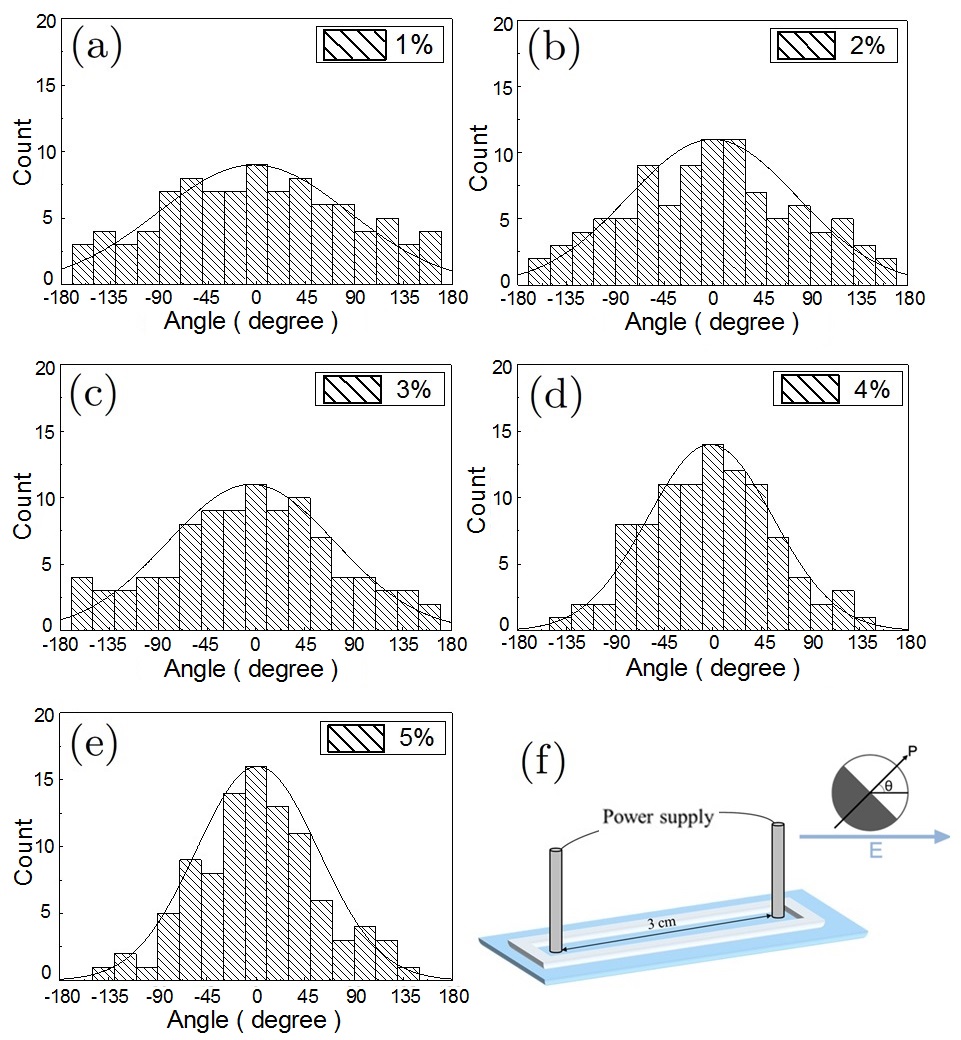}
\caption{(a)(b)(c)(d)(e) The plots show angular distribution in different concentration of H$_2$O$_2$ solution by count, where $\theta$ is the angle between the dipole of Janus particle and applied electric field. More concentrated the distribution to 0  $^{\circ}$ shows in higher concentration of H$_2$O$_2$ solution. (f) Device which provides DC electric field.}           

\label{induction loop} 
\end{center}
\end{figure}

To probe the electric properties on the surface in the asymmetric geometry during catalytic chemical reactions, we build a simple device which is dedicated to apply DC electric field to the channel as shown in Fig.5(f). The generated oxygen bubbles interference to electric field is avoided by opening the channel to the air. Assume excess electrons accumulate on Pt surface during the catalytic reactions, it will form an electric dipole which the dipole moment is subject to a torque when placing into an external electric field. So the dipole of Janus particle will be aligned toward the vector of external electric field. We observed the Janus particles in different concentrations of H$_2$O$_2$ solution in the channel and took the angular distribution of 100 Janus particles as shown in Fig.5. The higher concentration of H$_2$O$_2$ solution is, the more the angular distribution centralizes. It leads to the conclusion that generated dipole moment is related to the catalytic reaction of H$_2$O$_2$.

We report the self-propulsion of Pt-coated dielectric Janus particles can move in the reversed direction. Our results suggest that the motion is related to self-electrophoresis. The particles may move due to the electric field generated by itself. We figure out the direction of motion should be determined by the $\zeta$-potential of Janus particles and provide experimental measurements. Comparing with proposed mechanism of Pt-coated dielectric particles, we believe new measurement would lead new insights for this well known self-propulsion system. Although the detail processes on the chemical reaction are still not fully clear, results can be consist explained by a self-electrophoresis based mechanism. Self-electrophoresis mechanism is not yet be consider in dielectric based particles, our results may help people to reconsider self-generated electric field in conductive micro- and nano- active components. 
\\
\\
\\
\textit{Reference} \\
\text{[1]} Gangwal, S., Cayre, O. J., Bazant, M. Z., and Velev, O. D. (2008). Physical review letters, 100(5), 058302.\\
\text{[2]} Wurger, A. (2015). Physical review letters, 115(18), 188304. \\
\text{[3]} Jiang, H. R., Yoshinaga, N., and Sano, M. (2010). Physical review letters, 105(26), 268302.  \\
\text{[4]} Paxton, W. F., Sundararajan, S., Mallouk, T. E., and Sen, A. (2006). Angewandte Chemie International Edition, 45(33), 5420-5429.\\
\text{[5]} Mozaffari, A., Sharifi-Mood, N., Koplik, J., and Maldarelli, C. (2016). Physics of Fluids (1994-present), 28(5), 053107.\\
\text{[6]} Howse, J. R., Jones, R. A., Ryan, A. J., Gough, T., Vafabakhsh, R., and Golestanian, R. (2007). Physical review letters, 99(4), 048102.\\
\text{[7]} Ke, H., Ye, S., Carroll, R. L., and Showalter, K. (2010). The Journal of Physical Chemistry A, 114(17), 5462-5467.\\
\text{[8]} Lee, T. C., Alarcón-Correa, M., Miksch, C., Hahn, K., Gibbs, J. G., and Fischer, P. (2014). Nano letters, 14(5), 2407-2412.\\
\text{[9]} Paxton, W. F., Kistler, K. C., Olmeda, C. C., Sen, A., St. Angelo, S. K., Cao, Y., ... and Crespi, V. H. (2004). Journal of the American Chemical Society, 126(41), 13424-13431.\\
\text{[10]} Ebbens, S. J., and Howse, J. R. (2011). Langmuir, 27(20), 12293-12296.\\
\text{[11]} Wu, M., Zhang, H., Zheng, X., and Cui, H. (2014). AIP Advances, 4(3), 031326.\\
\text{[12]} Brown, A., and Poon, W. (2014). Ionic effects in self-propelled Pt-coated Janus swimmers. Soft matter, 10(22), 4016-4027.\\
\text{[13]} Gibbs, J. G., and Zhao, Y. P. (2009). Applied Physics Letters, 94(16).\\
\text{[14]} Wang, Y., Hernandez, R. M., Bartlett, D. J., Bingham, J. M., Kline, T. R., Sen, A., and Mallouk, T. E. (2006). Langmuir, 22(25), 10451-10456.\\
\text{[15]} Ebbens, S., Gregory, D. A., Dunderdale, G., Howse, J. R., Ibrahim, Y., Liverpool, T. B., and Golestanian, R. (2014). EPL (Europhysics Letters), 106(5), 58003.\\
\text{[16]} Hayase, K., and Hayano, S. (1978). Journal of Colloid and Interface Science, 63(3), 446-451.\\
\text{[17]} Domínguez, A., Fernández, A., González, N., Iglesias, E., and Montenegro, L. (1997). J. Chem. Educ, 74(10), 1227.\\
\text{[18]} Yamada, M., and Suzuki, S. (1984).  Analytical letters, 17(4), 251-263.\\
\text{[19]} Ghosh, S., and Moulik, S. P. (1998). Journal of colloid and interface science, 208(2), 357-366.\\
\text{[20]} Gittings, M. R., and Saville, D. A. (1998). Colloids and Surfaces A: Physicochemical and Engineering Aspects, 141(1), 111-117.\\
\text{[21]} Simmchen, J., Magdanz, V., Sanchez, S., Chokmaviroj, S., Ruiz-Molina, D., Baeza, A., and Schmidt, O. G. (2014). RSC advances, 4(39), 20334-20340.\\
\text{[22]} Esumi, K., Matoba, M., and Yamanaka, Y. (1996). Langmuir, 12(9), 2130-2135.\\
\text{[23]} Paxton, W. F., Baker, P. T., Kline, T. R., Wang, Y., Mallouk, T. E., and Sen, A. (2006). Journal of the American Chemical Society, 128(46), 14881-14888.\\
\text{[24]} Wang, Y., Hernandez, R. M., Bartlett, D. J., Bingham, J. M., Kline, T. R., Sen, A., and Mallouk, T. E. (2006). Langmuir, 22(25), 10451-10456.\\

\renewcommand\refname{Reference}
\bibliographystyle{plain}
\bibliography{jiang}

\end{document}